
\magnification=\magstep1
\font\medium=cmbx10 scaled \magstep1

\baselineskip=24pt
\parindent=20pt
\parskip=0pt
\topskip=0pt
\leftskip=0.0in
\rightskip=0.0in
\hsize=6.5 true in
\vsize=8.9 true in
\overfullrule=5pt
\hfuzz=1pt

\pageno=1
\vskip 3 true cm

{\medium \centerline {Landau Level Ground--State Degeneracy, and Its
 Relevance }}
{\medium \centerline {for a General Quantization Procedure}}

 \bigskip
 {\bf\centerline{Robert Alicki* }}
\bigskip

\centerline{Institute of Theoretical Physics and Astrophysics}
\centerline{University of Gda\'nsk, 80-952 Gda\'nsk, Poland}

\centerline{\bf John R. Klauder }
\bigskip
\centerline { Departments of Physics and Mathematics}
\centerline{ University of Florida, Gainesville, FL 32611}
\centerline {\bf  Jerzy Lewandowski** }
\bigskip
\centerline { Department of Physics }
\centerline{ University of Florida, Gainesville, FL 32611}
\bigskip\bigskip\bigskip\bigskip\bigskip\bigskip
\par
*) Kosciuszko Foundation Fellow at the Department of Physics, University of
 Florida, Gainesville.
\par
**) On leave of absence from the Institute of Theoretical Physics,
University of Warsaw, PL-00-681 Warsaw, Poland

\vfill\eject

\centerline{{\bf ABSTRACT}}
\par
The quantum dynamics of a two-dimensional charged spin $1/2$ particle is
 studied for general, symmetry--free curved surfaces and general,
 nonuniform magnetic fields that are, when different from zero, orthogonal
to the defining two surface. Although higher Landau levels generally lose
 their degeneracy under such general conditions, the lowest Landau level,
  the
ground state, remains degenerate. Previous discussions of this problem have
 had less
generality and/or used supersymmetry, or else have appealed to very general
 mathematical theorems from differential geometry. In contrast our
  discussion
relies on simple and standard quantum mechanical concepts.

\par
The mathematical similarity of the physical problem at hand and that of a
phase-space path integral quantization scheme of a general classical system
 is emphasized. Adopting this analogy in the general case leads to a
  general quantization procedure that is invariant under general coordinate
   transformations -- completely unlike any of the conventional
quantization
prescriptions -- and therefore generalizes the concept of quantization to
 new
and hitherto inaccesible situations.

\par
In a complementary fashion , the so-obtained picture of general
 quantization
helps to derive useful semiclassical formulas for the Hall current in the
case of a filling factor equal to one for a general surface and magnetic
 field.
\bigskip\bigskip
Short title: {\it Magnetic degeneracy and quantization}
\bigskip\bigskip
\centerline {PACS numbers:  03.65.Ca, 72.20.My}
\vfill\eject
{\bf 1. INTRODUCTION AND SUMMARY}
\par
For nonrelativistic electrons endowed with their usual spin magnetic moment
(i.e., $g_B=2$) motion in a two-dimensional plane perpendicular to a
 homogeneous
magnetic field has a number of interesting properties. Without taking the
 spin
contribution into account the energy levels of a free particle split into
 the
degenerate Landau levels endowed with the sequence of energy eigenvalues
$E_n=(n+{1\over 2})\hbar \omega_c, n=0,1,2,...,$ where $\omega_c=eB/mc$.
When the spin is included each level splits with half the states rising in
 energy and the other half falling in energy. Thanks to a proper magnetic
moment ($g_B=2$) those levels that rise exactly overlap with those levels
 that fall from the next higher Landau level leading to combined energy
  values
given by $E_n=n{\hbar}\omega_c, {n=0,1,2,...}$. While all levels but the
 lowest
contain spin up and spin down states the lowest level consists only of spin
down states and has exactly zero energy for any value of $\omega_c$. It is
 common to regard the level degeneracy as due to translational symmetry,
  and
for all but the lowest Landau level this viewpoint is correct. For the
 lowest
Landau level, however, an additional symmetry applies that preserves the
degeneracy even under circumstances where the degeneracy of the higher
levels is lifted. As we shall see the circumstances for which degeneracy of
the ground state remains are exceptionally broad including cases where the
magnetic field is not uniform in strengh as well as cases where a
 (non)uniform
magnetic field is everywhere orthogonal to a two surface that does not have
constant curvature. A surface of constant curvature such as the plane (zero
curvature) or the sphere (positive curvature) is necessary to have
degeneracy of the higher levels, but a generally symmetry-less surface --
loosely referred
to as a "potato", as may arise by deforming a sphere -- even in the presence
of a nonuniform magnetic field, maintains degeneracy of the lowest Landau
level.
\par
The existence of a degenerate ground state for electrons moving in the
presence
of nonuniform magnetic field everywhere perpendicular to a (compactified)
plane
has been known for some time$^1$; a compactified plane arises due to
periodic
boundary conditions, or, effectively, when the magnetic field vanishes
outside
some compact region. These properties have been demonstrated using
methods of supersymmetric quantum mechanics applied to underlying
 plane surfaces$^2$. Recently, the degeneracy of the ground state has
  been extended to cases of a nonuniform
magnetic field everywhere perpendicular to a general, compact,
symmetry-free underlying
surface$^3$. The methods entailed in this proof used contemporary
techniques in
differential geometry. In this paper we demonstrate that straightforward
techniques of nonrelativistic quantum mechanics are sufficient for this
more
general situation as well.
\par
{\bf 1.1 Euclidean Path Integral}
\par
Although our method of proof will involve partial differential equations,
we wish to present our basic results in the form of path integrals. The
purpose
behind this form of presentation is twofold: on one hand, path integrals
involve a functional formulation that is manifestly close in formal
appearence
to the underlying classical theory; and, on the other hand, the
ultimate
expressions may be given a form that makes manifest their covariance
under
coordinate transformations. This feature will be of considerable interest
when
attention is turned to a mathematical analog system, namely, that of
a phase-
space path integral for a general classical Hamiltonian which is at
once
rigorous in its formulation, and, simultaneously, covariant under
general
coordinate transformations. However, more on the analog system later (see
Sec.1.4).
\par
Consider, initially, a charged spin-1/2 particle moving on a two-dimensional
plane and subject to a uniform magnetic field perpendicular to the plane.
We assume also that the spin is polarized along the magnetic field. The
Euclidean space path integral for the propagator is given, in a convenient
gauge, by the formal expression ($\hbar =1, g_B=2$)

$${\cal N }\int{ \exp {\Bigl\{i\int{(m{\omega_c}y\dot x )dt}\ -\
{1\over 2}\int{[ m({\dot x}^2 + {\dot y}^2)-{\omega_c }]dt}
\Bigr\}}
 {\Pi} dx dy }\ .\eqno(1.1)$$
The propagator represents the matrix element

$$<x'',y''|e^{-HT}|x',y'> \ , \eqno(1.2)$$
where $H$ has a spectrum given by $n\omega_c , n=0,1,2,...$. Let us next take
the limit $\omega_c \to\infty $; theoretically we can do so by letting
$m\to 0$, while empirically such a limit is approached by choosing
large magnetic fields. The result of such a limit is the matrix elements
of a {\it projection operator},

$$\lim_{m\to 0}<x'',y''|e^{-{\cal }HT}|x',y'>\ =\ <x'',y''|\Pi |x',y'> \
.\eqno(1.3)$$
In the present case the explicit form is easily worked out, and one finds
that
$$<x'',y''|\Pi |x',y'>\ =\ (eB/2{\pi}c)\exp \bigl\{{i\over 2}(eB/c)(y''+y')
(x''-x')$$
$$-{1\over 4}(eB/c)[(y''-y')^2 + (x''-x')^2]\bigr\}\ .\eqno(1.4)$$
It is readily verified that this expression represents the integral kernel
of a projection operator. The rank of the projection operator $\Pi$ -- which
equals the degeneracy of the lowest Landau level -- is given in turn by
$\int <x,y|\Pi |x,y> dxdy$, which diverges in the present case.
\par
It is also useful to consider the matrix elements of the projection operator
somewhat more abstractly. To this end we introduce the notation ${\cal K }
(x'',
y'';x',y')$ instead of $<x'',y''|\Pi |x',y'>$, and observe that for ${\cal K}$
to
represent a projection operator it is necessary and sufficient that
${\cal K}^{\ast}(x'',y'';x',y')={\cal K} (x',y';x'',y'')$ and
${\cal K} (x''',y''';x',y')=\int{ {\cal K} (x''',y''';x'',y''){\cal K}
 (x'',y'';x'y') dx''dy''}$.
When these conditions are satisfied then the rank of the so-determined
 projection
operator is given by $\int {\cal K }(x,y;x,y)dxdy$.
\par
Let us generalize our physical situation so that the electron moves in the
presence of a local potential $V(x,y)$ as well as the uniform magnetic field.
However, we do not Euclideanize the potential, only the kinetic term, so the
expression of interest is represented by the formal path integral
$${\cal N }\int{ \exp {\Bigl\{i\int{[m{\omega_c}y\dot x - V(x,y)]dt}\ -\
{1\over 2}\int{[ m({\dot x}^2 + {\dot y}^2)-{\omega_c }]}dt\Bigr\}}
 {\Pi} dx dy }\ .\eqno(1.5)$$
In the limit that $m\to 0$ we still expect that the Hilbert space
collapses
to the lowest Landau level, but in general the result is no longer a
 projection
operator. Instead, there is a dynamical evolution generated by the
 hermitian
Hamiltonian which is determined by an integral kernel that is given
by$^4$
$${\cal H} (x'',y'';x',y')=\int{ {\cal K} (x'',y'';x,y)V(x,y){\cal K}
 (x,y;x',y')
 dxdy}\ . \eqno(1.6)$$
In words, the Hamiltonian is given by the two-sided projection of the
 potential
$V$ onto the lowest Landau level. We denote the ultimate limit (a unitary
propagator for the lowest Landau level at zero mass or high magnetic field
 limit) by
$$ K(x'',y'',t'';x',y',t')\ =\ <x'',y''|e^{-i{\cal H}T}|x',y'>\ =$$
$$\lim_{m\to 0}{\cal N }\int{ \exp {\Bigl\{i\int{[m{\omega_c}y\dot x -
V(x,y)]dt}\ -\ {1\over 2}\int{[m({\dot x}^2 + {\dot y}^2)-{\omega_c }]dt}
\Bigr\}} {\Pi} dx dy }\ ,\eqno(1.7)$$
where $T=t''-t'>0$. Additionally, it follows that
$$\lim_{t''\to t'}K(x'',y'',t'';x',y',t')\ =\ {\cal K}(x'',y'';x',y')\
.\eqno(1.8)$$
\par
Why have we chosen to rotate only the kinetic energy and not the potential
energy to imaginary time? The answer lies in our desire to obtain a genuine
Wiener measure on $(x,y)$ path space so as to put the path integral
expression
for a unitary time evolution in the projected Hilbert subspace on a sound
 mathematical foundation.
In particular, we note that
$$K(x'',y'',t'';x',y',t')=$$
$$\lim_{m\to 0}(2\pi c/eB)\int \exp\bigl\{i(eB/c)
\int ydx\ -\ i\int V(x,y)dt\bigr\}\exp \bigl\{\int (\omega_c /2)
\bigr\}
d\mu_W (x,y)\ , \eqno(1.9)$$
where $\mu_W$ denotes a pinned Wiener measure as commonly appears in the
 Feynman-Kac formula. The expression $\int ydx$ is to be interpreted as a
(Stratonovich) stochastic integral, in which case this path integral
expression for $K$ is without any ambiguity and rigorously defined
for each
 $m>0$; convergence as $m\to 0$ is assured for a wide class of potentials.
 $^4$
  As
a well-defined integral one may also consider its rigorous reformulation
under coordinate transformations. Under such transformations the phase
 factor
 transforms under the rules of the ordinary calculus in spite of the fact
  that
the functions involved are Brownian and not classical (e.g., $C^1$) in
character; these transformation properties are the result of the
Stratonovich
(mid-point) prescription. The Brownian motion itself transforms as one
 might
 expect:
as initially formulated the two-dimensional, planar Brownian motion was
described by Cartesian coordinates; after the transformation the same
two-dimensional, planar Brownian motion should be described, in general,
by curvilinear coordinates.
\vfill\eject
\par
{\bf 1.2 General Field and Surface}
\par
With the foregoing elementary and familiar problem as background we turn
 our attention to present analogous results in more general circumstances.
  For
present purposes we introduce intrinsic coordinates $x^1$ and $x^2$ lying
 in the surface, which as usual, may be described by a Riemanian metric
$ds^2=g_{ab}(x)dx^adx^b$ . The surface may be compact or noncompact and
 may have an arbitrary genus (number of handles), although for the most
  part we restrict attention to a simply connected manifold. In addition,
   we assume
there is a magnetic field present that is described by a vector potential
$A_b(x)$ in the standard way, $ B_{ab}(x)=\partial_a A_b(x)-
\partial_bA_a(x)$.
As an antisymmetric tensor in two dimensions it is clear that  $B_{ab}(x)=
\epsilon _{ab} \lambda(x)$, where $\epsilon_{ab}$ is the Levi-Civita tensor
density and $\lambda (x)$ is a scalar density. Without loss of generality,
 we shall always orient the surface so that the total magnetic flux is
 nonnegative.
\par
The path integral that represents the desired generalization of the ones
given earlier reads
$$\lim_{m\to 0} {\cal N}\int \exp \bigl\{i(e/c) \int A_b(x)dx^b\ -\
i\int V(x) dt\bigr\}$$
$$\times \exp \bigl\{-{1\over 2}m\int g_{ab}(x){\dot x}^a{\dot x}^b dt\
 +\
{e\over {2mc}}\int s^{ab}(x)B_{ab}(x)dt\bigr\}{\Pi} {\sqrt {g(x)}}dx^1
dx^2\ ,\eqno(1.10)$$
where the spin tensor $s^{ab}=\sqrt {g}{\epsilon}^{ab}/2$.
The structure of this expression has been chosen with several issues in
 mind.
The terms in the exponent, except the one containing $V(x)$, plus the form
of the integration  measure describe the Euclidean propagator of a charged
spin-$1/2$ particle moving on the curved surface in the presence of a
 magnetic
field everywhere orthogonal to the surface. In particular, the final
 term in
the exponent represents a generalization of the term $\int (\omega_c/2)
dt$
and describes the interaction of the polarized spin-$1/2$ with the
 magnetic
 field ($g_B=2$). For this form of interaction the degeneracy of the
lowest
  Landau level is not destroyed by a nonuniform field and/or a curved
   geometry.
Moreover the energy of the lowest Landau level remains equal to zero.
 These
 facts lie at the heart of what is proved in the following Section.
As a consequence, when $V\equiv 0$ and in the limit $m\to 0$ the path
 integral
 (1.10) leads to an integral kernel for a projection operator on a
degenerate lowest Landau level, while for $V\not=0$ a unitary evolution
on the corresponding Hilbert subspace is obtained.
\par
There are two kinds of transformations of the formal path integral of
interest.
 By construction the expression is invariant under coordinate
transformations, $x\to {\bar x} ={\bar x} (x)$, assuming that the
indicated
quantities transform like tensors of the appropriate kind. A second
 kind of
 transformation involves
a change of gauge of the vector potential, $A_b(x)\to A_b(x)+
\partial_b
\Lambda (x)$. The only consequence of such a transformation is
 the appearence
of a total derivative leading to a phase factor of the form
$\exp\{i(e/c)
[\Lambda(x'')-\Lambda(x')]\}$. Such a factor only affects the
 local phase
of the wave function, a modification without physical content.
Of course,
transformations that combine both gauge and coordinate changes
 are important
as well as we shall see in the next subsection.
\par
{\bf 1.3 Reinterpretation in Phase Space}
\par
It is often useful to take the mathematical formulation appropriate
 to one
physical situation and reinterpret it in an entirely different physical
situation. Hamiltonian mechanics for particles and ray optics provides
just one example of the utility of such a reinterpretation. Quantization
of two-dimensional particles in a magnetic field and a phase-space path
integral quantization of a particle, as we now shall see, provides yet
another example.
\par
We return, first of all, to the case of a particle moving on the plane in
the presence of a uniform magnetic field and an auxiliary potential $V$.
For present purposes let us introduce new variables, viz.,
$$ q\ =\ \sqrt{eB/{\Omega} c}\ x\ \ \ ,\ p\ =\ \sqrt{eB{\Omega}/c}\ y\ ,$$
$$\nu \ =\ eB/mc\ =\ \omega_c\ ,\ h(p,q)\ =\ V(x,y)\ .\eqno(1.11)$$
In terms of these variables the former path integral, Eq.(1.5), assumes
the form
$$\lim_{\nu\to\infty}{\cal N}\int \exp \bigl\{i\int [p{\dot q}-h(p,q)]dt
\ -\ {1\over {2\nu}}\int ({\Omega}^{-1}{\dot p}^2+\Omega {\dot q}^2-
 \nu^2)dt\bigr\}
{\Pi} dp dq\ .\eqno(1.12)$$
Apart from the limit and the $\nu$-dependent factor in the integrand the
expression in question resembles a formal phase-space path integral.
The additional factor may be interpreted as a regularizing factor, more
specifically as a continuous-time regularization, for in the limit
$\nu\to\infty $, the factor in question formally becomes unity. To gain
insights into the consequences of such a regularization we first specialize
to the case $h=0$, and define
$${\cal K}(p'',q'';p',q')\ =\ \lim_{\nu\to\infty}{\cal N}\int \exp \bigl
\{i\int p{\dot q}dt
\ -\ {1\over {2\nu}}\int ({\Omega}^{-1}{\dot p}^2+\Omega {\dot q}^2-\nu^2)
dt\bigr\}
{\Pi} dp dq$$
$$ =\ \exp \bigl\{{i\over 2}(p''+p')(q''-q')-{1\over 4}[{\Omega}^{-1}
(p''-p')^2+
{\Omega} (q''-q')^2]\bigr\}\ ,\eqno(1.13)$$
as follows from (1.4),  with the proviso that we have rescaled the
 integration
measure to absorb the prefactor, namely, $(eB/2{\pi}c)dxdy=dpdq/2{\pi}$.
 It
readily follows that
$${\cal K}(p''',q''';\- p',q')=\int {\cal K}(p''',q''';p'',q''){\cal K}
(p'',q'';p',q') dp''dq''/2{\pi}\eqno(1.14)$$ and
${\cal K}^{\ast}(p'',q'';p',q')={\cal K}(p',q';p'',q'')$; therefore
 ${\cal K}$
represents a projection operator, but a projection onto what? Just as
 in the
 planar motion in a magnetic field, the projection operator projects
 onto the
 relevant Hilbert space for the subsequent quantum mechanics. In the
  present
  case ${\cal K}$ denotes a projection operator on $L^2({\bf R}^2,
   dpdq/2{\pi})$
onto the relevant functional Hilbert space for the problem at hand.
 Nevertheless
the integral kernel for the projection operator is, at first sight,
 unfamiliar in
 its quantum mechanical meaning. Insight into that meaning
is gained by first observing that ${\cal K}$ is a positive definite
 function,
i.e., satisfies
$$\sum {\alpha}^{\ast}_j{\alpha}_k{\cal K}(p_j,q_j;p_k,q_k)\ =\
\int |\sum {\alpha}_k {\cal K}(p,q;p_k,q_k)|^2 dpdq/2{\pi}\ \geq 0\ ,
\eqno(1.15)$$
in virtue of the properties of ${\cal K}$ previously given.
As a consequence the (Gel'fand, Naimark, Segal) GNS Theorem$^5$ asserts
 that there
exists a representation of ${\cal K}$ as the inner product of two
Hilbert space
 vectors that is unique
up to unitary equivalence; namely, there exists an abstract Hilbert space
${\bf H}$ and vectors $|p,q>\in {\bf H}$, for all $(p,q)\in {\bf R}^2$,
 such
that ${\cal K}(p'',q'';p',q')=$  $<p'',q''|p',q'>$ for all argument
pairs. In
special cases -- such as the one presently under consideration -- these
 vectors
are generated by a transitively acting group (or a group up to factor) on
a fixed fiducial vector, but this situation is far more the exception
 than the rule.
\par
In the present case the appropriate states are given by
$$|p,q>\ =\ e^{-iqP}e^{ipQ}|\Omega> \eqno(1.16)$$
for all $(p,q)\in {\bf R}^2$, where $Q$ and $P$ denote irreducible
 self-adjoint
 Heisenberg operators and $|\Omega >$ is a normalized vector that
satisfies
 $(\Omega Q+iP)|\Omega>=0$. In terms of the Schr\"odinger
 representation it
 follows that
$$<p'',q''|p',q'>\ =\ <\Omega|e^{-ip''Q}e^{i(q''-q')P}e^{ip'Q}
|\Omega>$$
$$=\ {\sqrt {\Omega /\pi}}\int \exp\{-{\Omega}x^2/2-ip''x+ip'
(x+q''-q')\}
\exp\{-{\Omega}(x+q''-q')^2/2\} dx$$
$$=\  \exp \bigl\{{i\over 2}(p''+p')(q''-q')-{1\over 4}[{\Omega}^{-1}
(p''-p')^2
+{\Omega}(q''-q')^2]\bigr\}$$
$$\equiv {\cal K}(p'',q'';p',q')\ .
\eqno(1.17)$$
The GNS Theorem then effectively asserts the unique association of
 the Weyl
group and the
Heisenberg operators with this particular kernel. Of course, the
 states
 $|p,q>$
in question are just the familiar canonical coherent states$^6$,
 which in
 ${\bf H}$
 admit a resolution of unity in the form
$${\bf 1}\ =\ \int |p,q><p,q| dpdq/2{\pi}\ .\eqno(1.18)$$
These states provide a representation basis for an arbitrary vector
 $|\psi>
\in {\bf H}$, given by $\psi (p,q)\equiv<p,q|\psi>$, with an inner
 product
given by $\|\psi\|^2\,\equiv \,\int|\psi (p,q)|^2dpdq/2{\pi}\,= $
 $<\psi |\psi >$.
Finally, the propagator that arises when $h(p,q)\not=0$ is just the
coherent-state
 matrix element of the evolution operator, namely
$$<p'',q''|e^{-i{\cal H}T}|p',q'>$$
$$=\ \lim_{\nu\to\infty}{\cal N}\int \exp \bigl\{i\int
 [p{\dot q}-h(p,q)]dt
\ -\ {1\over {2\nu}}\int ({\Omega}^{-1}{\dot p}^2+\Omega
{\dot q}^2-\nu^2)dt\bigr\}
{\Pi} dp dq $$
$$\equiv \ K(p'',q'',t'';p',q',t')\ .\eqno(1.19)$$
In this expression
$${\cal H}(p'',q'';p',q')\equiv <p'',q''|{\cal H}|p',q'>\ =\
\int <p'',q''|p,q>h(p,q)<p,q|p',q'> dpdq/2{\pi}\ ,\eqno(1.20)$$
or abstractly
$${\cal H}\ =\ \int h(p,q) |p,q><p,q| dpdq/2{\pi}\ ,\eqno(1.21)$$
which relates the Hamiltonian operator ${\cal H}$ and its c-number
representative
$h(p,q)$.

Let us interpret the integral in (1.19) as one involving a Wiener
measure and a
Stratonovich stochastic integral. In that case it becomes
appropriate to discuss
 coordinate transformations. In particular consider a change of
 canonical
  coordinates $\bar p=\bar p(p,q),\ \bar q=\bar q(p,q)$ for
  which $pdq=
\bar p d\bar q +dF(\bar q,q)$. This equation which holds for
 classical ($C^1$)
 functions holds for Brownian paths as well. In light of the
 discussion in
 the previous subsection, we have chosen to link a gauge
 transformation with
  a suitable coordinate transformation so as to preserve the
  form of the
   classical action (and of the associated classical equations
    of motion). With the Wiener measure reinterpreted as planar
     Brownian motion expressed in curvilinear coordinates, an
     expression such as (1.19) transforms covariantly under a
     canonical change of coordinates.

\par
{\bf Recapitulation}
\par
In this subsection we have reinterpreted the mathematics appropriate
to a charged
spin-$1/2$ particle moving in a two-dimensional plane in the presence
 of
a uniform magnetic field and an auxiliary potential as a phase-space,
 path-
integral quantization  procedure. Admittedly the reinterpreted expression
has the form of a phase-space path integral apart from the unusual
$\nu$-dependent
factor in the integrand. This factor has apparently introduced a metric
 into phase space for the purpose of quantization where none seems to be
  present
in alternative quantization procedures, e.g., the standard Schr\"odinger
 prescription.
However, we assert that a metric is implicitly used in Schr\"odinger
quantization when one recognizes that the Schr\"odinger rules of
quantization
work correctly only in certain coordinates, namely Cartesian coordinates.
$^7$
A flat metric appears when it is recognized that global Cartesian
coordinates exist only in a globally flat space.
The role of the $\nu$-dependent factor is, of course, to regularize
 the formal
integral which then may be reinterpreted as a well-defined Brownian
 motion
integral.
\par
Based on an analogous reinterpretation of the motion of charged spin-$1/2$
 particles in a magnetic field we shall, in the next subsection, propose a
 quantization scheme for phase spaces endowed with general symplectic forms
and general and unrelated metric structures. In so doing we will encounter
an unexpected surprise related to the quantization of such systems, namely,
each path does {\it not} contribute to the path integral with equal weight
in the general case.
\par
{\bf 1.4  Quantization of General Systems}
\par
As was the case in the previous subsection we initiate our discusion with
the kinematics. Let the phase-space variables be denoted by
$\xi=(\xi^1,\xi^2)$
, set $(e/c)A_a=a_a , (e/c)B_{ab}=b_{ab}$, $m=1/{\nu}$, in which case
attention focusses on
$$\lim_{\nu\to \infty} {\cal N}\int \exp \bigl\{i \int a_b(\xi)
d\xi^b\bigr\}$$
$$\times \exp \bigl\{-{1\over {2\nu}}\int g_{ab}(\xi){\dot \xi}^a
{\dot \xi}^b
 dt\ +\
{\nu \over 2}\int s^{ab}(\xi )b_{ab}(\xi )dt\bigr\}{\Pi} {\sqrt
 {g(\xi )}}
d{\xi}^1
d{\xi}^2\ .\eqno(1.22)$$
In the next Section we shall prove that the kernel defined by this
expression corresponds to a projection operator on a nontrivial
 subspace
 of the Hilbert
space $L^2(\Gamma ,{\sqrt g }d\xi^1 d\xi^2)$, ($\Gamma $ denotes the
phase-space
manifold). This subspace will be identified with the Hilbert space
${\bf H}$ of the quantum system. As a consequence the kernel satisfies
$${\cal K}(\xi '';\xi ')\ =\ \int {\cal K}(\xi '';\xi){\cal K}(\xi;\xi ')
{\sqrt g} d\xi^1 d\xi^2\ ,$$
$${\cal K}^{\ast}(\xi '';\xi ')\ =\ {\cal K}(\xi ';\xi '')\ .
\eqno(1.23)$$
We shall analyse  phase-spaces with an ${\bf R}^2$ topology and
derive the formula for the dimension of ${\bf H}$, and even a
local expression for
the semiclassical density of quantum states. The case of a
compact Riemanian
phase-space manifold will be also discussed and illustrated
by examples.
For the latter case the compactibility condition
$${1\over 2}\int b_{ab}(\xi )d\xi^a\wedge d\xi^b \ =\ 2\pi n,\ \ \ \
 n=1,2,3,...\eqno(1.24)$$
should hold; the dimension of ${\bf H}$ is then finite and given by
$$ D\ =\ n\ +\ 1\ -\ g\ ,\eqno(1.25)$$
where $g$ denotes the number of handles on the surface.
\par
 From the viewpoint of classical mechanics $a_b(\xi )d\xi^b$ denotes
the one form whose exterior derivative
$$da_b (\xi )d\xi^b\ =\ \partial_aa_b(\xi )d\xi^a \wedge d\xi^b\ =\
 {1\over 2}
b_{ab}(\xi )d\xi^a \wedge d\xi^b \eqno(1.26)$$
denotes the symplectic two form on the manifold. In simple cases,
 namely
canonical coordinates, the one form is just $pdq$ and the symplectic
 form
then is $dp\wedge dq$ . The symplectic form is, in this simple case,
 the same
volume element that appears in the formal path integral measure,
 namely $\Pi
dpdq$ . It is noteworthy in the general case that the volume element
required in the path integral measure is not (proportional to) the
symplectic
form volume element. In the general case the volume element
$\sqrt{g(\xi )}
d\xi^1d\xi^2$ appears in the path integral measure while the
 symplectic
form is given by $ {1 \over 2}b_{ab}(\xi )d\xi^a \wedge d\xi^b.$
  This fact flies in the face of conventional wisdom that in a
  path integral "all paths enter
with equal weight". Of course, the additional weighting factor
$\exp [(\nu /2)\int s^{ab}b_{ab}(\xi )dt]$ belies this conventional
 wisdom as well.
\par
Conventionally, the phrase "symplectic form`` is reserved to a
 nondegenerate
skew-symmetric matrix, $\omega_{ab}$, such that $\omega_{ab}\omega^{bc}
=-\delta _a^c$. In this paper, however, we refer loosely to the
skew-symmetric matrix $b_{ab}$ as a symplectic form even if it may
 be degenerate in some regions and even when it is not degenerate
 it may fail to be a square root of unity in the sense noted above.
 Our justification for this terminology arises from the fact that
 $b_{ab}$ are the coefficients in the exterior derivative of the
 one form $a_b (\xi)d\xi^b$ that figures in the action functional
 for the system at hand$^8$.
\par
The kernel ${\cal K}(\xi '';\xi ')$ is a positive-definite functional,
 and as such, according to the GNS Theorem, may be represented as the
 inner product
of (not necessarily normalized) vectors $|\xi >\equiv |\xi^1,\xi^2>$
in an abstract Hilbert space ${\bf H}$
, namely,
        $$<\xi ''|\xi '>\ =\ {\cal K}(\xi '';\xi ')\ .\eqno(1.27)$$
These vectors are continuously labelled and, in virtue of the projection
property of ${\cal K}$ , they admit a resolution of unity in ${\bf H}$
according to
$${\bf 1}\ =\ \int |\xi ><\xi | \sqrt{g} d\xi^1d\xi^2\ .\eqno(1.28)$$
These are just the properties that make the vectors $\{ |\xi >\}$ into
a set
of coherent states. It must be emphasized, however, that in the general
 case
there is no few-parameter unitary representation of a group (or a group
up to the factor) that generates all the states $|\xi >$ as unitary
transformations
of a fixed fiducial vector. However convenient such a group may be there
is, in
the general case, no symmetry of the phase-space manifold that would
support
the existence of such a transitively acting group. The difference in
 viewpoint
regarding quantization advocated here could not be greater than the
conventional quantization viewpoint in which one
promotes several of the classical phase-space variables to self-adjoint
 operators
appropriate to some low-dimensional closed Lie algebra. These two
quantization
procedures coincide
for a limited number of cases, but will surely lead to different
results in
the general case. The existence of the physical analog of the
quantum Hall
effect speaks to the validity of the alternative quantization scheme
advocated
in this subsection in the general case.
\par
The introduction of a nonvanishing Hamiltonian and nontrivial dynamics
proceeds as in the elementary case. The propagator is given by
$$K(\xi '',t'';\xi ',t')
\ =\ \lim_{\nu\to \infty} {\cal N}\int \exp \bigl\{ i \int [ a_b(\xi)
{\dot{\xi}}^b
-h(\xi )] dt\bigr\}$$
$$\times \exp \bigl\{-{1\over {2\nu}}\int g_{ab}(\xi){\dot \xi}^a{\dot
 \xi}^b dt\ +\
{\nu\over 2} \int s^{ab}(\xi )b_{ab}(\xi )dt\bigr\}{\Pi} {\sqrt {g(\xi )}}
d{\xi}^1 d{\xi}^2$$
$$\equiv <\xi ''|e^{-i{\cal H}T}|\xi '>\ .\eqno(1.29)$$
Here ${\cal H}$ and $h$ are related by
$${\cal H}\ =\ \int h(\xi )|\xi ><\xi | \sqrt{g} d\xi^1d\xi^2\
 .\eqno(1.30)$$
To ensure that a unitary evolution exists it is sufficient for
 ${\cal H}$
to be essentially self-adjoint on the finite linear span of the
 coherent
states.
\par
With the final formulas we have achieved our goal of presenting a
manifestly
coordinate invariant quantization procedure appropriate to a general
 symplectic
form and geometry of the underlying two manifold. One should mention
 that the present approach to quantization has been extended to
  K\"ahler
  manifolds
of an arbitrary even dimension,$^9$ and for flat phase-spaces Wiener
 measure in
(1.9) may be replaced by a probabilistic measure for a general Poisson
 process.$^{10}$
\vfill\eject
\par{\bf 2. THE STRUCTURE OF THE LOWEST LANDAU LEVEL}
\par
Consider an electron moving on an arbitrary smooth two-dimensional
surface $\Gamma$ as described in
Sec. 1.2. The path integral expression (1.10) with a fixed value
of the
mass parameter $m$ and with $V(x)\equiv 0$ yields the integral kernel
of the operator $\exp \{-H[A,g]T\}$ where ($\hbar = 1$)
$$ H[A,g]\ =\ -{1\over 2m}\bigl[{1\over {\sqrt
g}}(\partial_a+i(e/c)A_a)
g^{ab}{\sqrt g}(\partial_b+i(e/c)A_b) \bigr]\ -\ {e\over {2mc\sqrt{g}}}
B_{12}
\ . \eqno(2.1)$$
It follows from (2.1) that the limit $m\to 0$ in the path integral
(1.10)
for $V \equiv 0$ is equivalent to taking the following operator limit
(in the sense of matrix elements)
$$ \Pi \ =\ \lim_{T\to \infty}\exp\{-H[A,g]T\}\ .\eqno(2.2)$$
The limit operator $\Pi$ exists and is a nontrivial projection
operator
in the Hilbert space $L^2(\Gamma, {\sqrt g}dx^1dx^2)$
if and only if : $H[A,g]\geq 0$ and there exists a nontrivial
subspace
${\bf H}$ of normalizable eigenvectors $\phi $ satisfying
$$ H[A,g]\phi \ =\ 0\ .\eqno(2.3)$$
In the following we shall construct the solutions of Eq. (2.3) for a
manifold
$\Gamma$ admiting a global parametrization (${\bf R}^2$ topology)
generalizing
the Aharonov-Casher$^1$  approach to a flat surface with an arbitrary
magnetic
field, and then we shall briefly discuss two examples of compact
manifolds.
Before doing this we should take advantage of the fact that for any
two-dimensional surface one can always choose a (local) coordinate
system, say $u$ and $v$, $(u,v)\in {\bf R}^2$, such that the metric becomes
conformally flat i.e.$^{11}$
$$ds^2\ =\ e^{2w(u,v)}\bigl(du^2\ +\ dv^2\bigr)\ .\eqno(2.4)$$
In this special coordinate system the matrix elements of the
Hamiltonian
$H[A,g]$ are given by the following expression
$$<\psi |H[A,g]|\phi>\ =\ -{1\over {2m}}\int
{\psi}^{\ast}e^{-2w}\Big\{\bigl[{\partial_u}+i(e/c)A_u)\bigr]^2\phi
+
\bigl[{\partial_v}+i(e/c)A_v) \bigr]^2\phi\Big\}e^{2w}dudv $$
$$ -\ {e\over {2mc}}\int {\psi}^{\ast}(\partial_u A_v-\partial_v A_u)\phi
\,dudv $$
$$\equiv \ {1\over 2m}\int dudv ({\cal D}\psi )^{\ast}{\cal
D}\phi\eqno(2.5)$$
where
$$ {\cal D}\phi \ =\ \big[(\partial_u - i\partial_v)
+i(e/c)(A_u-iA_v)\big]
\phi \ .\eqno(2.6)$$
 From (2.5) and (2.6) it follows that $H[A,g]\geq 0$ indeed and that
the ground states ( with polarized spin) are all solutions of the
following
 equation
 $$\big[(\partial_u - i\partial_v) +i(e/c)(A_u-iA_v)\big]
\phi \ =\ 0\ .\eqno(2.7)$$
Obviously the relevant solutions must be square integrable with
respect to
the measure ${\sqrt g}dx^1dx^2$ and must satisfy the topological
constraints in the case of compact manifold $\Gamma$. Equation (2.7)
 gives
us control on the singularities
of $\phi$. Indeed, in the neighborhood of any point there always exists
a local non-singular
solution, say $\rho$, which does not vanish.$^{13}$ Any other solution
 $\phi$
can be
expressed in terms of $\rho$ as $\phi(u,v)\ =\ f(u - iv) \rho (u,v)$,
 $f$ being a holomorphic function. Hence,  any singularity (or zero)
of $\phi$ is a singularity (zero) of a holomorphic function. We
conclude
from this that a square integrable solution is supposed to be smooth.
This
has implications on the topological restrictions. Mathematically,
Eq. (2.7) defines a holomorphic bundle and $\phi$ is a global
section. There are known strong mathematical methods which give us
the dimension of the space of solutions to (2.7) in the compact case in
terms of
topological invariants: the flux of  the magnetic field and the Euler
characteristic of the surface. We shall illustrate them in Sec. 2.2. On
 the other hand, in Sec. 2.1 we show that even in a non-compact,
  topologically flat, case the magnetic flux and the integral of a
  Gauss curvature -- provided that they are finite -- determine the
   dimension of ${\bf H}$.

{\bf 2.1 Surface with $R^2$ Topology}
\par
We assume now that there exists a global coordinate system ($u,v$)
satisfying
(2.4). Then it follows from Eq.(2.7) that the
subspace ${\bf H}$ of the
ground states is spanned by the linearly independent functions
$$\phi_k (u,v)\ =\ (u-iv)^ke^{-F(u,v)}e^{iG(u,v)}\eqno(2.8)$$
with $k=0,1,2...,N(=D-1)\leq \infty$, and  real functions $F,G$
satisfying
the equations
$$ (\partial^2_u+\partial^2_v)F(u,v)\ =\
(e/c)(\partial_uA_v-\partial_vA_u)
\ ,\eqno(2.9)$$
$$ (\partial^2_u+\partial^2_v)G(u,v)\ =\
(e/c)(\partial_uA_u+\partial_vA_v)
\ .\eqno(2.10)$$
The condition of square integrability of $\phi_k$ demands that the
function
$(u^2+v^2)^k\exp$  $[-2F(u,v)+2w(u,v)]$ should decay at least as
$(u^2+v^2)^{-(1+\epsilon )}$ for $|u|,|v|\to \infty $ with $\epsilon
>0$. Suppose now that the following integrals are finite $(\Phi \geq 0)$
$$\Phi \ =\ (e/c)\int (\partial_uA_v-\partial_vA_u)dudv\
,\eqno(2.11)$$
$$\Psi \ =\ -\int (\partial^2_u+\partial^2_v)w(u,v)dudv \eqno(2.12)$$
The solution of (2.9) can be written as
$$F(u,v)\ =\ {e\over {4\pi c}}\int du'dv'\bigl\{[\partial_uA_v(u',v')-
\partial_vA_u(u',v')]\ln [(u-u')^2+(v-v')^2]
\bigr\}\ .\eqno(2.13)$$
For large $|u|^2+|v|^2$ we obtain the following estimation, using (2.9),
(2.11),(2.12) and (2.13),
$$|\phi_k (u,v)|^2 \exp [-2w(u,v)] \sim (u^2+v^2)^{k-(\Phi
+\Psi)/2\pi }\ .\eqno(2.15)$$
Hence to attain square integrability $k$ must satisfy the inequality
$$k\ <\ {1\over 2\pi }(\Phi +\Psi )\ -\ 1\ .\eqno(2.16)$$
The expressions for  $\Phi $ and $\Psi $ can be easily transformed
into a geometric, coordinate independent form
$$ \Phi \ =\ {e\over {2c}}\int (\partial_a A_b(x)-\partial_b A_a(x))
dx^a\wedge dx^b\ ,
\eqno(2.17)$$
$$ \Psi \ =\ {1\over 2}\int R(x) \sqrt{g(x)} dx^1dx^2 \eqno(2.18)$$
where $R$ is the scalar curvature given by the Riemann tensor of $g$
$$ R\ =\ {R^{\alpha\beta}}_{\alpha\beta} =
-2e^{-2w}(\partial^2_u+\partial^2_v)w\ .\eqno(2.19)$$
We emphasise however, that in this case  $\Psi$ and $\Phi$ are not
topological invariants.
Finally, from (2.16) our expression for the dimension of the lowest
Landau level reads
$$D\ =\ largest\ integer\ less\ than
\Bigl[  {1\over {2\pi }}\Phi\ +\ {1\over {2\pi} }\Psi \ -\ 1
\Bigr]\  .\eqno(2.20)$$
\par
Clearly for infinite $\Phi$ and/or $\Psi$ the subspace ${\bf H}$ is
infinite dimensional. However,even in this case the following formula for
the semiclassical density of electronic states on the surface $\Gamma
$ (with a unidirectional magnetic field)is valid as can be seen from Eqs.
 (2.17),(2.18) and (2.20)
$$dN(x)\ =\ {e\over {4\pi c}}(\partial_a A_b(x)\ -\ \partial_b
A_a(x)) dx^a\wedge dx^b\ +\ {1\over {4\pi }} R(x) \sqrt{g(x)}
dx^1dx^2\ .\eqno(2.21)$$
\vfill\eject
\par
{\bf 2.2 Compact surfaces}
\par
The case of a compact two-dimensional manifold $\Gamma$ with an
arbitrary
genus $ g = 0,1,2,...,$ can be discussed using geometrical methods. First
 of all the vector potential $A_b$ and
the coordinates at which
the metric tensor $g_{ab}$ takes the form (2.4) are defined only
locally and subject to a suitable gauge/coordinate transformation
from a one to another local domain.
The (normalized) integrals ${1 \over 2\pi}\Phi$, the magnetic charge,
and ${1 \over 2\pi}\Psi$, the Euler characteristic, are now
topological invariants and can take only integer values, namely
$${1 \over 2\pi} \Phi \ =\ n,\ \ \ n=0,1,2,...\eqno(2.22)$$
$${1 \over 2\pi}\Psi \ =\ 2(1-g),\ \ \ g=0,1,2,...\ .\eqno(2.23)$$
The condition (2.22) is the famous Dirac condition on the monopole
while the condition (2.23) is the Gauss-Bonnet Theorem.$^{11}$
As mentioned in Section 1.4 the Riemann-Roch-Hirzebruch-Atiyah-Singer
(see for example Ref.12) index  theorem gives the
dimension $D$ of the lowest Landau level as
$$ D\ =\ n+(1-g)\ =\ \big[{1\over {2\pi }}(\Phi + \Psi)-1\big]+g\
\eqno(2.24)$$
if $n\ >\ 2-2g$ or $g=0$ and when $n\ \le \ 0$, necessarily
$D\ =\ 0$ .
Note that Eq. (2.24) extends the formula (2.20) to compact manifolds.
Here, again, in the semiclassical limit ($n\gg 1+g$, unidirectional
 magnetic
 field) the local
expression
(2.21) for the density of states remains valid.
The manifest expressions for the wave functions which span ${\bf
H}$
in the case of compact $\Gamma$ are obtained as the solutions of
Eq. (2.7)
which satisfy the topological constraints. For the sake of
illustration
we present three particular examples.
\vfill\eject
\par
{\bf Example 1. Potato}
\par
We consider here Eq. (2.7) on a 2-surface $\Gamma$ which is
topologically equivalent to a sphere. The genus $g=0$, now, and we know
from the classification of Riemann surfaces that $\Gamma$ is
conformal to a sphere equipped with the natural metric. The  coordinates
$(u,v)$ cannot be extended to the entire surface $\Gamma$. However, in
 this case,
there exist  `spherical' coordinates
 $(\theta, \alpha)$ such that the scalar
product $ds^2 = g_{\mu\nu} dx^\mu dx^\nu , (\mu ,\nu = \theta ,\alpha)$
takes on the following appearance
$$ds^2 = e^{2w}(d\theta^2 + sin^2\theta d\alpha^2)\ .\eqno(2.25)$$
Let $A_\mu$ be a vector potential carrying the magnetic charge $\Phi =
2\pi n$.
According to (2.24), the number of linearly independent solutions of
(2.7) is
$$D\ =\ n+1\ .\eqno(2.26)$$ We shall derive them below, but first here
is an
outline of our strategy.
We write $A$ as
$$A_\mu \ =\ n {\tilde A}_\mu + a_\mu\eqno(2.27)$$
where ${\tilde A}$ is a vector potential of the uniform magnetic field
corresponding to a magnetic charge $n_0 =1$. Next, we solve Eq. (2.7) with
${\tilde A}_\mu $ and $a_\mu$, respectively,  substituted for $A_\mu$. In
the first case we find
two linearly independent solutions, ${\tilde \psi}_{(1)}$ and ${\tilde
 \psi}_{(2)}$,
and in the second case a single solution denoted by $\phi'$. This is
consistent with (2.26).
Finally, we define wave functions $\phi_{(1)},...,\phi_{(n+1)}$ by
$$\phi_{(i+1)}\ =\ \phi'({\tilde \psi}_{(1)})^{i}({\tilde \psi}_{(2)})
^{n+1-i}\
 .\eqno(2.28)$$
Every $\phi_{(i)}$ is a solution to (2.7) with the vector potential
(2.27). It is also easy enough to see (details below)  that the
$\phi_{(i)}$-s are linearly independent, hence they form a basis of the
solutions.
More specifically, to express the vector potential  $A$ we divide
$\Gamma$ onto  two hemispheres and on each of them fix a gauge (if
$n>0$ then there is no global gauge on $\Gamma$).
Then $A$ and  an associated wave function $\phi$ may be written as
$$(A_\mu,\ \phi) = \cases {(A_\mu^+,\ \phi^+)& if $\theta \le {\pi
\over 2}+\epsilon$;
\cr (A_\mu^-,\ \phi^-)& if $\theta \ge {\pi \over 2} -
\epsilon$}\eqno(2.29)$$
where $A_\mu^\pm$ and $\phi^\pm$ are well defined on the hemispheres,
and on the intersection of the two hemispheres we glue them by a gauge
transformation
$$A^+ = A^- + nd\alpha,\ \ \ \phi^+ = e^{-in\alpha}\phi^-\ .\eqno(2.30)$$
Through $n$ in the exponent, the gauge transformation contains the
information about the magnetic charge.
For the uniform magnetic field we choose a vector potential
$${\tilde A}^\pm = {1 \over 2}(\pm 1 + cos\theta)d\phi \ .\eqno(2.31)$$
The solutions  corresponding to ${\tilde A}$ have the following form
$$\phi_{(1)} \ \ = \ \  \cases {cos{1 \over 2}\theta&,
\cr e^{i\alpha}cos{1 \over 2}\theta&;}\eqno (2.32)$$
$$\phi_{(2)} \ \ = \ \  \cases {e^{-i\alpha}sin{1 \over 2}\theta&,
\cr sin{1 \over 2}\theta&.}$$
On the other hand, the term $a_\mu$ in (2.27) is a globally defined
covariant vector field.
It follows from the fact that $\Gamma$ is simply connected, that
 $a_\mu$ can
be decomposed into the form
$$a_\mu\ \ = \ \ {\partial_{\mu}G} +
{\epsilon_{\mu}}^{\nu}{\partial_{\nu}b}\eqno(2.33)$$
with $G$ and $b$ being real functions on $\Gamma$. The solution of (2.7)
corresponding to $a_\mu$ is
$$\phi' \ \ =\ \ e^{-(b+iG)}\ .\eqno(2.34)$$

We have learned from this example that for a simply connected surface
it is enough to find ground states for a uniform magnetic field
which has the flux $2\pi$ and for all the magnetic fields of zero
flux.
Then, ground states for an arbitrary magnetic field are generated
algebraically
 from the previous ones.
\par

{\bf Example 2. Donut}
\par
We consider here a surface topologically equivalent to a torus. This
means that
 the
genus $g=1$,  and the dimension of the space of solutions to (2.7) given by
(2.24) becomes $D=n$. Geometry of the surface is, up to a pointwise
dependent rescaling, equivalent to the geometry of the quotient: the
plane $R^2$ equipped with the flat metric $du^2 + dv^2$ divided by the
group of translations generated by two vectors
$$X\ =\ (2\pi, 0),\ \ \ \ V\ =\ (u_0, v_0),\ \ \ v_0\ >\ 0\ .\eqno(2.35)$$
The topological conditions  which have to be satisfied by a wave
function $\phi$ of a particle interacting with a vector potential $A_\mu$,
which
 has
the topological charge $n$, take the  form of  certain  periodicity
conditions. They can be written as
$$A_\mu(u+2\pi,v)dx^\mu\ \ = \ \ A_\mu(u,v)dx^\mu + {2\pi n\over
v_0}dv,$$
$$\phi(u+2\pi,v)\ \ =\ \exp \bigl(-{2\pi n i \over v_0}v\bigr)\phi(u,v),
\eqno(2.36)$$
$$A_\mu(u+u_0,v+v_0)dx^\mu\ \ = \ \ A_\mu(u,v)dx^\mu,$$
$$\phi(u+u_0,v+v_0)\ \ =\ \ \phi(u,v),$$
where $(x^\mu)=(u,v)$.
To solve  Eq. (2.7) we shall use the same trick as in the previous
example.
We decompose $A$ into the  sum of a vector potential of a uniform
magnetic field which carries the  topological charge and the rest.
However, in the case of a sphere, magnetic field determined uniquely a
gauge class of corresponding vector potentials. Now, this one to one
correspondence
does not hold. The ambiguity
consists of magnetic `vacua' given by
constant vector potentials of the form
$$A'_u = \beta_u,\ \ \ A'_v = \beta_v\ ,\eqno(2.37)$$
$\beta$s being constant numbers.  In other words, we write $A$ as
$$A_\mu\ \ =\ \ n{\tilde A}_\mu  +  a_\mu,\eqno(2.38)$$
where for ${\tilde A}$ we can choose
$${\tilde A}_\mu dx^\mu\ \ =\ \ {v_0u-u_0v \over v_0^2}dv\ ;\eqno(2.39)$$
but unlike in (2.33), the Hodge decomposition of $a_\mu$ reads
$$a_\mu\ \ = \ \ \beta_\mu + {\partial_{\mu}G} +
{\epsilon_{\mu}}^{\nu}{\partial_{\nu}b},\eqno(2.40)$$
where $\beta_u$ and $\beta_v$ are real  constants.  After the substitution
of (2.38) and (2.40)
into (2.7), the second and the third term of $a_\mu$ [see the RHS of
(2.40)]
can be eliminated from  Eq (2.7) in the same way as in Example 1,
i.e., by
introducing $\psi$ such that
$$\phi \ \ = \ \ \psi  e^{-(b+iG)}\ .\eqno(2.41)$$
In  that way,  we are left with the following equation
$$({\partial \over \partial u} -i {\partial \over \partial v}\ + \ n
 {v_0u-u_0v \over 2 v_0^2}\ +\
\beta)\psi = 0.\eqno(2.42)$$
 The general solution to (2.42) which  satisfies
the first periodicity condition, i.e., with respect to the
translations generated by the vector $(2\pi ,0)$, can be expressed as
$$\psi \ = l({\bar z})\ \exp {\Bigl( n {{iu_0v^2 }\over{2 v_0^2}}}  +
n {{v^2-2iuv}\over {2{v_0}}} + 2i\beta v\Bigr)\ ,\eqno(2.43)$$
where $z:=u+iv$, $\beta:=\beta_u+i\beta_v$ and the function $l$ is periodic
 with respect to the vector $(2\pi,
0)$,
$$l(\bar z) = \sum_{k=-\infty}^{\infty}a_k e^{ik{\bar z}}\ .\eqno(2.44)$$
Applying the second periodicity condition, that with respect to
$z\rightarrow
z + \beta$, we obtain the condition
$$a_{k+n} = a_k{\exp}\Bigl[-i(k+{n\over 2})(u_0-iv_0) + 2iv_0\beta\Bigr]\ .\
\eqno(2.45)$$
Hence, we can fix $n$ arbitrary values for $a_0,...,a_{n-1}$ and
determine
by (2.45) all other $a_k$. It is easy to see that (2.45) guaranties that the
 obtained sum which gives $l$  converges for every $z$ since $v_0>0$.

Summarising this example, we could see above the mechanism which determines
the number of independent (polarized spin) ground states as determined by the
 magnetic flux.

\par
{\bf Example 3. Arbitrary surface but topologically trivial magnetic
field}
\par
Mathematically, Eq (2.7) defines a holomorhic bundle, and a global
solution forms a holomorphic section.  However, in the previous
 examples we
did not necessarily have  to apply the theory of holomorphic bundles.
 We could
 just explicitly derive the solutions.  On the other hand, if the
  surface has
 higher genus then  straightforward computations would be  very
  complicated
  and we only have the formula (2.22). Now, we would like to
  concentrate on
   the case when the magnetic flux vanishes, i.e.,
when $n=0$.   We shall present now
how the mathematics works for this example.
Suppose a wave function $\phi$ is a solution to (2.7). We shall see
that
there are no other solutions linearly independent of $\phi$. Indeed,
suppose
that $\phi'$ also solves (2.7) with the same vector potential $A$.
Then
necessarily
$$\phi' = f({\bar z})\phi\eqno(2.46)$$
where $f({\bar z})$ is an anti-holomorphic function of $z$. The only
(anti)holomorphic functions on a compact surface are constant functions.
However, if $\phi$
has a zero in some point then perhaps $f$ can have a pole which is
compensated
by $\phi$. Therefore, we have to study zeros of $\phi$, and here
mathematics gives us a precise answer. First, as we mentioned before every
 zero of $\phi$ is a zero of a holomorphic function.
Hence, any such zero is of the kind $({\bar z} - {\bar z_0})^p$. Second, we
have a formula which expresses the magnetic flux
by the zeros of $\phi$ and their orders; this expression reads
$$2\pi n \ \ =\ \ \Phi = 2\pi \sum p_i \eqno(2.47)$$
summed over all zeros. But in our case  $\Phi=0$ it follows that every order
$p_i\ \ =\ \ 0$. Hence, $\phi$ {\it cannot vanish at any point}.

Summarising, we have seen that, if $n=0$  there are two possibilities: either
 there exists exactly
one solution or none. A solution exists if and only if $A$ can be
written in the Landau gauge in the form
$$A_a(x) = \sqrt{g(x)}{\epsilon_a}^c(x) {\partial_c}b(x)\eqno(2.48)$$
with $b$ being a global real function on the surface.
\par
{\bf 2.3  Application to Quantization of General Systems}
\par
The results of the previous sections have immediate application to the
 problem
of quantization of a general system discussed in Section 1.4. Namely
 treating
now the two dimensional surface $\Gamma$ as a phase space of a certain
physical system we obtain the representation of the Hilbert space
${\bf H}$ of
 the quantized system. ${\bf H}$ is identified with subspace of the
 Hilbert
space $L^2(\Gamma, {\sqrt g}d\xi^1d\xi^2)$ which contains functions
satisfying the polarization condition
$$ \Bigl\{\bigl[{1\over {\sqrt g}}(\partial_a+ia_a)
g^{ab}{\sqrt g}(\partial_b+ia_b) \bigr]\ +\ s^{ab}b_{ab}\Bigr\}\phi
\ =\ 0
\ . \eqno(2.49)$$
In the special coordinate system the solutions of (2.49 ) are given
 by the
 solution
of Eq.(2.7) with $(e/c)A_b$, $(e/c)B_{ab}$ replaced by $a_b$,
 $b_{ab}$ and
 in different topological cases we proceed as in Sections 2.1, 2.2.
 Having
  found the solutions $\phi_k$ of Eq.(2.7),
which span ${\bf H}$, we may construct the reproducing kernel
 ${\cal K}$ as
$${\cal K}(\xi'';\xi')\ =\ \sum \beta_{kl}\phi_k(\xi'')
\phi_l^{\ast}(\xi')\
 ,\eqno(2.50)$$
where $\beta_{kl}$ are coefficients of a matrix inverse to the
 Gramm matrix
with coefficients $\alpha_{kl}=\int \phi_k^{\ast}(\xi )\phi_l
(\xi ){\sqrt
 {g(\xi )}}d\xi^1d\xi^2$,
and then complete the quantization scheme presented in Section 1.4.
\par
{\bf 2.4 Quantum Hall Current}

The motion of an electron in a general magnetic field and on an
arbitrary
 surface $\Gamma$ as discussed in Section 1.2 is described by the
  propagator
  (1.10).
However, according to the reinterpretation given in Sections 1.3
 and 1.4
this propagator
may be treated as a quantum propagator for the classical system
 with a
 phase-space $\Gamma$ and an action functional
$${\cal A}\ =\ \int \bigl[{e\over c}A_b(x){\dot x}^b-V(x)\bigr]dt\ .
\eqno(2.51 )$$
The corresponding Euler-Lagrange equations read
$${e\over c}B_{ab}(x){\dot x}^b\ =\ \partial_a V(x)\ .\eqno(2.52)$$
Consider now two points $x'$ and $x''$ on $\Gamma$ connected by a
 curve $C$.
The total electric current $J_C$ which flows through the curve
 $C$ for
 the case of
fully occupied first Landau level may be calculated using the
following
semiclassical arguments. Let us treat the electrons as a fluid
 with the
 local
surface density given by  Eq.(2.21) and the local velocity
 ${\dot x}^a$
 which satisfies Hamiltonian equation (2.52). First from Eq.
 (2.21)we
  obtain
$$dN(x)\ =\ {e\over {4\pi c}}(\partial_a A_b(x)\ -\ \partial_b
A_a(x)) dx^a\wedge dx^b\ +\ {1\over {4\pi }} R(x) \sqrt{g(x)}
dx^1dx^2\ $$
$$= \bigl[{e\over {4\pi c}}(\partial_a A_b(x)\ -\ \partial_b
A_a(x)) \ +\ {1\over {8\pi }} R(x) \sqrt{g(x)}\epsilon_{ab}
\bigr] dx^a\wedge dx^b\ .\eqno(2.53)$$
 Then using Eqs.(2.52),(2.53) and the fact that $B_{ab} = ({\sqrt
  {B_{rs}B^{rs}g/2}})\epsilon_{ab}$ we have
 $$J_C\ =\ {e^2\over {2\pi c}}\int_C B_{ab}{\dot x}^adx^b
 +{1\over {4\pi }}\int_C R(x)\sqrt {g(x)}{\epsilon}_{ab}{\dot x}^a
 dx^b$$
 $$ =\ {e\over{2\pi c}}[V(x'')-V(x')]
 + {1\over {4\pi }}\int_C {R(x)\over {\sqrt {B_{ab}B^{ab}/2}}}dV(x)\
  \eqno(2. 54)$$
The first term on the RHS of Eq.(2.54) gives the standard expression
 for the quantum Hall current with the filling factor equal to
 one$^{14}$
 while the second
one is a geometric correction due to the curvature. For a flat
 surface
and uniform magnetic field the standard expression is verified
experimentally
with an
amazing accuracy. We expect also that the generalized formula (2.54)
is applicable far beyond the semiclassical limit for physically
interesting cases (here semiclassical regime
corresponds to the case where the typical magnetic length
$[eB/c]^{-1/2}$ is
much smaller then the other relevant length scales). This wider
 applicability
 is due to the fact that
all quantum corrections which can be derived from the expansion in
 path integral
(1.10) around the classical trajectory effectively cancel
in the integral along the curve $C$ as long as the external
potential $V(x)$
varies very slowly at the ends $x',x''$. One should notice that
 Eq. (2.54) makes sense for $B_{ab}\not= 0$, which is in agreement
  with the semiclassical picture. For an application of the above
  results to the description of an anomalous Hall current due to
   the anomalous magnetic moment of an electron
in the case of flat surface but general magnetic field see Ref.15.

\goodbreak
{\bf Acknowledgements}
\par
 RA is grateful to the Department of Physics, University of
 Florida, for
hospitality. The research of RA was  partially supported   by
 the Polish
 Comitee for Scientific Research (KBN project PB 1436/2/91).
 JL was
  supported from National Science Foundation contracts PHY
 9107007 and
partially by the Polish Comitee for Scientific Research (KBN)
 through grant
2 0430 9101.
\bigskip\bigskip
{\bf References}
\nobreak
\par
\item{1. } Y. Aharonov and A. Casher,Phys.Rev.{\bf A19}, 2461 (1979);
  R. Jackiw,  Phys.Rev.{\bf D29},  2375  (1984).
\item{2. }L.E.  Gendenshtein, Sov.J.Nucl.Phys. {\bf 41},166 (1985);
L. E. Gendenshtein and I. V. Krive, Sov.Phys.Usp. {\bf 28}, 645 (1985).
\item{3. }P. Maraner, Mod.Phys.    Lett.{\bf A7}, 2555 (1992).
\item{4. }J.R. Klauder and I. Daubechies, Phys.Rev.Lett.{\bf 52}, 1161
 (1984)
; I. Daubechies and J.R. Klauder, J.Math.Phys.{\bf 26}, 2239 (1985).
\item{5. }G.G. Emch, {\it Mathematical and Conceptual Foundations of 20th-
Century Physics} (North-Holland, Amsterdam, 1984).
\item{6. }J.R. Klauder and B.-S. Skagerstam, {\it Coherent States:
 Applications in Physics and Mathematical Physics} (World Scientific,
  Singapore, 1985).
\item{7. }P.A.M. Dirac, {\it The Principles of Quantum Mechanics. Third
 Edition}
 (Oxford 1947; p. 114).
\item{8. }L. Faddeev and R. Jackiw, Phys.Rev.Lett.{\bf 60}, 1692 (1988).
\item{9. }J.R. Klauder and E. Onofri, Int.J.Mod.Phys.{\bf A4}, 3930 (1989).
\item{10.}R. Alicki and J.R. Klauder, J.Math.Phys. {\bf 34}, 3867 (1993).
\item{11.}H.M. Farkas and I. Kra, {\it Riemann Surfaces} (Springer-Verlag,
 N.York, 1980).
\item{12.}R.C. Gunning, {\it Lectures on Riemann Surfaces} (Princeton Univ.
 Press, 1966).
\item{13.}S.K. Donaldson and P.B. Kronheimer, {\it The Geometry of Four
 Manifolds} (Oxford Univ. Press, 1991)
\item{14.} R.E. Prange and S.M. Girvin (Eds.) {\it The Quantum Hall
 Effect,
Second Edition} (Springer-Verlag, N.York, 1990).
\item{15.} R. Alicki and J.R. Klauder, {\it Anomalous Quantum Hall
 Current
Driven
\item{  }by a Nonuniform Magnetic Field}, preprint 1993.

\bye